\documentclass[12pt,preprint]{aastex}
\usepackage{graphicx}
\usepackage{natbib}
\usepackage{amssymb,amsmath}
\usepackage{cite}

\shorttitle{GRB120729A: External Shock for Prompt Emission and Afterglow}
\begin{document}
\title{GRB120729A: EXTERNAL SHOCK ORIGIN FOR BOTH THE PROMPT GAMMA-RAY EMISSION AND AFTERGLOW}
\author{Li-Ye Huang\altaffilmark{1,2}, Xiang-Gao Wang\altaffilmark{1,2,3,5}, WeiKang Zheng\altaffilmark{3}, En-Wei Liang \altaffilmark{1,2}, Da-bin Lin\altaffilmark{1,2},Shi-Qing Zhong\altaffilmark{1,2}, Hai-Ming Zhang\altaffilmark{1,2},Xiao-Li Huang\altaffilmark{1,2}, Alexei V. Filippenko\altaffilmark{3,4}, Bing Zhang\altaffilmark{5}}

  \altaffiltext{1}{GXU-NAOC Center for Astrophysics and Space Sciences, Department of Physics, Guangxi University, Nanning 530004, China; wangxg@gxu.edu.cn, lew@gxu.edu.cn}
  \altaffiltext{2}{Guangxi Key Laboratory for the Relativistic Astrophysics, Nanning 530004, China}
  \altaffiltext{3} {Department of Astronomy, University of California, Berkeley, CA, 94720-3411, USA; weikang@berkeley.edu}
  \altaffiltext{4} {Miller Senior Fellow, Miller Institute for Basic Research in Science, University of California, Berkeley, CA 94720, USA; afilippenko@berkeley.edu}
  \altaffiltext{5}{Department of Physics and Astronomy, University of Nevada, Las Vegas, NV 89154, USA; zhang@physics.unlv.edu}

\begin{abstract}

Gamma-ray burst (GRB) 120729A was detected by \emph{Swift}/BAT and {\it Fermi}/GBM, and then rapidly observed by {\it Swift}/XRT, {\it Swift}/UVOT, and ground-based telescopes. It had a single long and smooth $\gamma$-ray emission pulse, which extends continuously to the X-rays. We report Lick/KAIT observations of the source, and make temporal and spectral joint fits of the multiwavelength light curves of GRB 120729A. It exhibits achromatic light-curve behavior, consistent with the predictions of the external shock model. The light curves are decomposed into four typical phases: onset bump (Phase I), normal decay (Phase II), shallow decay (Phase III), and post-jet break (Phase IV). The spectral energy distribution (SED) evolves from prompt $\gamma$-ray emission to the afterglow with photon index from $\Gamma_{\rm \gamma}=1.36$ to $\Gamma \approx 1.75$. There is no obvious evolution of the SED during the afterglow. The multiwavelength light curves from $\gamma$-ray to optical can be well modeled with an external shock by considering energy injection, and a time-dependent microphysics model with $\epsilon_B\propto t^{\alpha_B}$ for the emission at early times, $T < T_{\rm 0} + 157$~s. Therefore, we conclude that both the prompt $\gamma$-ray emission and afterglow of GRB 120729A have the same external shock physical origin. Our model indicates that the $\epsilon_B$ evolution can be described as a broken power-law function with $\alpha_{\rm B,1} = 0.18 \pm 0.04$ and $\alpha_{\rm B,2} = 0.84 \pm 0.04$. We also systematically investigate single-pulse GRBs in the \emph{Swift} era, finding that only a small fraction of GRBs (GRBs 120729A, 051111, and 070318) are likely to originate from an external shock for both the prompt $\gamma$-ray emission and afterglow.

\end{abstract}


\section{INTRODUCTION} \label{sec:Introduction}

Gamma-ray bursts (GRBs) are extremely energetic transient events in the Universe, signifying catastrophic events that involve core collapse in some massive stars and mergers of two compact objects \citep[e.g.,][]{kumarzhang15}. These events power an energetic, relativistic jet, which has powerful $\gamma$-ray emission with an isotropic equivalent energy $\sim 10^{50}$--$10^{55}$ erg. Observationally, the prompt $\gamma$-ray emission lasts from milliseconds to several thousand seconds, with most of the light curves showing rapid variability. Following the prompt $\gamma$-ray emission, the blast wave interacts with the circumburst medium and produces an afterglow, which is in principle detectable at X-ray through radio wavelengths.

The most popular model for GRBs is the standard fireball model, in which the broadband afterglow is from external shocks \citep[][]{meszarosrees97,sari98}, while the prompt $\gamma$-ray emission is from internal shocks \citep[][]{rees98,kobayashipiran97,daigne98}. Recent studies suggest that the observed afterglow data can be quite consistent with the predictions of the external shock model \citep[e.g.,][]{wang15}. Although the prompt emission is observed much earlier, it is less understood than the afterglow \citep{zhang11}. Alternative models have been proposed suggesting that the prompt emission may be from the photosphere \citep[][]{paczynski86,goodman86,meszarosree00,meszaros02,peer08}, magnetic dissipation regions \citep[][]{lyutikov03,zhangyan11}, or external shocks \citep[][]{reesmesazros92,meszarosrees93,dermer99}.

The rapid variability poses a problem for the external shock model. However, observationally there do exist some GRBs that only have a single smooth peak, or just a few long and temporally separated peaks with a large single smooth peak. The simple smooth profiles could arise from an external shock \citep[e.g.,][]{mcmahon04,dermer04,ramirez07,guidorzi07,golkhou14,golkhou15,burgess16}.

GRB 120729A is an interesting case with a single smooth pulse of $\gamma$-ray emission. Furthermore, the prompt $\gamma$-ray emission detected by \emph{Swift}/BAT smoothly extends to X-rays. Our observations are presented in \S 2. In \S 3, we perform the temporal and spectral analyses, and suggest the same physical origin (an external shock) for both the prompt $\gamma$-ray emission and afterglow. Then, in \S 4, we model the light curves with an external shock model by considering energy injection and time-dependent microphysics. Our discussion and conclusions are given in \S 5.  Temporal and spectral slopes are defined as $F\propto t^{-\alpha} \nu^{-\beta}$ throughout this paper.

\section{OBSERVATIONS}

GRB 120729A was triggered by the \emph{Swift}/BAT on 2012 July 29 (UT dates are adopted) at $10:56:14$, with $T_{90} = 71.5$~s \citep[][]{ukwatta12a, palmer12}. It was also detected by the {\it Fermi}/GBM with $T_{90} = 25$~s \citep {rau12}. The discrepancy of the detected duration ($T_{\rm 90}$) is caused by the difference in sensitivity of the BAT and GBM energy bands \citep{qiny13}. The X-ray Telescope (XRT) and the UV-optical Telescope (UVOT) onboard \emph{Swift} began observing the X-ray and optical afterglows of GRB 120729A at 68~s and 77~s after the BAT trigger, respectively \citep{ukwatta12b, oates12}.

\begin{figure}[htbp]
\centering
\includegraphics[angle=0,width=0.8\textwidth]{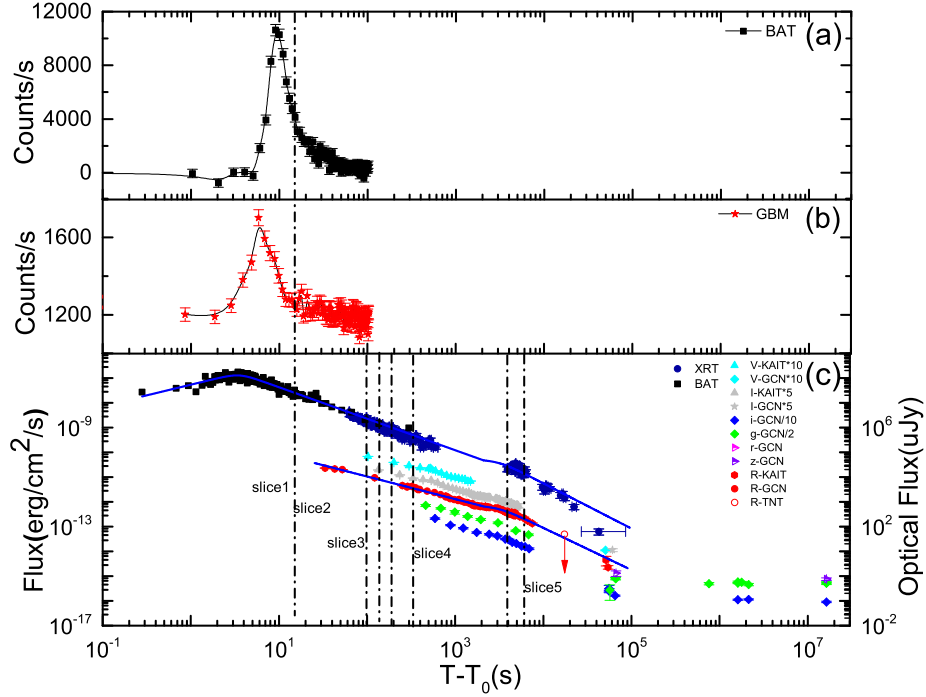}
\caption{Multiwavelength light curves of the prompt $\gamma$-ray emission and afterglow of GRB 120729A. (a) The prompt $\gamma$-ray emission detected by {\em Swift}/BAT, with $T_{90}=71.5$ s; (a) the prompt $\gamma$-ray emission detected by {\it Fermi}/GBM, with $T_{90}=25$ s; (c) multiwavelength light curves of the afterglow, and the {\it Swift}/BAT data converted to 0.3--10 keV, presented together. Note that {\em Swift}/BAT prompt-emission data extend to the X-rays smoothly. The vertical dashed lines mark the time slices of interest for the afterglow spectral analysis. The ground-based optical telescopes rapidly followed up the {\em Swift} trigger; e.g., KAIT and RAPTOR responded to the {\em Swift} trigger 98~s and 27.9~s after the burst, respectively. Some data are taken from \citet{cano14}.}
\label{Gam-opt}
\end{figure}

We downloaded the BAT data from the NASA {\em Swift} website, and used a Python source package  {\it gtBurst}\footnote{https://github.com/giacomov/gtburst} to extract light curves and spectra. The {\em Swift}/XRT light curve and spectrum are taken from the {\it Swift} Burst Analyzer (Evans et al. 2010)\footnote{http://www.swift.ac.uk/burst\_analyser/00599037/}. We also downloaded the {\it Fermi}/GBM data of GRB\,120729A from the {\it Fermi} Archive website\footnote{ftp://legacy.gsfc.nasa.gov/Fermi/data/}. We extract the light curve and spectrum from the {\it Fermi}/GBM data with our Python code. Spectral fitting package {\it Xspec} was used for our spectral analysis. Both the BAT and GBM light curves exhibit a single smooth pulse (Figure \ref{Gam-opt}a,b), and the BAT data smoothly connect to the XRT data (Figure \ref{Gam-opt}c).

GRB 120729A was rapidly followed up by ground-based telescopes: robotic 2~m Faulkes Telescope North (FTN) \citep{virgili12}, 2~m Liverpool Telescope (LT), 0.82~m Instituto de Astrofisica de Canarias IAC80 telescope (IAC), 3.6~m Telescopio Nazionale Galileo (TNG), 10.4~m Gran Telescopio Canarias (GTC), Wide Field Camera (WFCAM) on the United Kingdom Infrared Telescope (UKIRT) \citep{im12}, 0.4~m Rapid Telescopes for Optical Response (RAPTOR) \citep{wren12}, Virtual Telescope \citep{masi12}, 1.5~m Russian-Turkish telescope (RTT150) \citep{khamitov12}, 1.5~m Observatorio de Sierra Nevada (OSN) \citep{gorosabel12}, Burst Optical Observer and Transient Exploring System \citep{gorosabel12}, James Clerk Maxwell Telescope (JCMT) \citep{smith12}, and 0.8~m Tsinghua University -- National Astronomical Observatory Telescope (TNT) \citep{xin12}.

The 0.76~m Katzman Automatic Imaging Telescope (KAIT; \citep[][]{filippenko01,liw03}) at Lick Observatory also responded automatically to the {\it Swift}/BAT trigger and began imaging the field 98.0~s later. In this paper, we report the KAIT follow-up observations, which were performed in the $B$, $V$, $R$, and $I$ filters. Data reduction was carried out following standard routines in IRAF\footnote{IRAF is distributed by the National Optical Astronomy Observatory (NOAO), which is operated by AURA, Inc., under a cooperative agreement with the NSF.} package. The photometry is reported in Table \ref{table:magnitude}. Figure \ref{Gam-opt}(c) shows the optical light curves, which include the data from KAIT and Cano et al. (2014). The redshift of GRB 120729A was measured to be $z = 0.80$ \citep{tanvir12}. Note that BAT detections start 3.08~s before the BAT trigger (which occurred near the peak); thus, we shift $T_0$ to be this time of initial detection.

\section{MULTIWAVELENGTH LIGHT-CURVE BEHAVIOR}

An empirical model with a broken power-law function was employed to fit the light curves for temporal analysis \citep[e.g.,][]{liang07,lil12,wang15}:
\begin{equation} F=F_{\rm 0}\left [
\left (   \frac{t}{t_{\rm b}}\right)^{\omega\alpha_{\rm 1}}+\left (
\frac{t}{t_{\rm b}}\right)^{\omega\alpha_{\rm 2}}\right]^{{\rm -1}/\omega},
\end{equation}
where $\alpha_{\rm 1}$ ($\alpha_{\rm 2}$) is the temporal slope before (after) the break time $t_{\rm b}$ and $\omega$ represents the sharpness of the break. We found that the multiwavelength light curves can be well decomposed into four phases, as shown in Figure \ref{Gam-opt}(c). The BAT light curve smoothly onsets with a slope of $\alpha_{\rm BAT,I}=-0.86\pm0.11$ (Phase I) and peaks at $\sim T_0+3.5$~s. The flux smoothly connects to the X-ray band and decays as a power law from $T_0+3.5$~s to $T_0+1950$~s (Phase II), with the power-law index $\alpha_{\rm BAT,II}=\alpha_{\rm X,II}=1.27\pm0.09$ for both the BAT and  X-ray bands. The $R$-band light curve also decays as a power law with $\alpha_{\rm R,II}=0.96\pm0.02$ in Phase II. Subsequently, the flux remains almost constant in both the X-ray and $R$ bands from $T_0+1950$~s to $T_0+3153$~s (Phase III) with $\alpha_{\rm X,III}=0.18$ (fixed) and $\alpha_{\rm R,III}=0.18\pm0.12$, and then transits to a post-jet break (Phase IV) with $\alpha_{\rm X,IV}=1.83$ (fixed) and $\alpha_{\rm R,IV}=1.70\pm0.09$.

To get more information, the analysis of broadband spectral energy distributions (SEDs) needed to be done. The optical data were corrected for Galactic extinction based on the burst direction \citep[][]{schlegel98}, with $A_V=0.444$~mag, $A_R=0.351$~mag, $A_I=0.244$~mag, $A_g=0.534$~mag, and $A_i=0.275$~mag. The Galactic hydrogen column density in the burst direction is $N_{\rm H}=2.15 \times 10^{\rm 21}$ cm$^{\rm -2}$ (Willingale et al. 2013). The extinction law of the host galaxy was taken to be that of the Small Magellanic Cloud (SMC; $R_V=2.93$). The line-of-sight value of $N_{\rm H}$ in the host galaxy is $\sim 1.0\times 10^{\rm 21}$~cm$^{\rm -2}$, derived from the time-integrated X-ray afterglow spectrum and fixed at this value in our time-resolved spectral fits. The ``Xspec'' package was employed for the spectral analysis. We subdivided the broadband data into five temporal ranges (as marked in Figure \ref{Gam-opt}). Slice 1 covers the single pulse of $\gamma$-ray emission from BAT to GBM in the time interval $T_0$ + [0,~15]~s. We take Slices 2, 3, and 4 to cover both the $\gamma$-ray (BAT) and early afterglow (X-ray and optical) emission in respective time intervals $T_0$ + [15,~95]~s, $T_0$ + [95,~135]~s, and $T_0$ + [185,~330]~s. Slice 5 covers the late afterglow in the time interval $T_0$ + [3900,~6050]~s.

\begin{figure}[htbp]
\includegraphics[angle=0,width=0.5\textwidth]{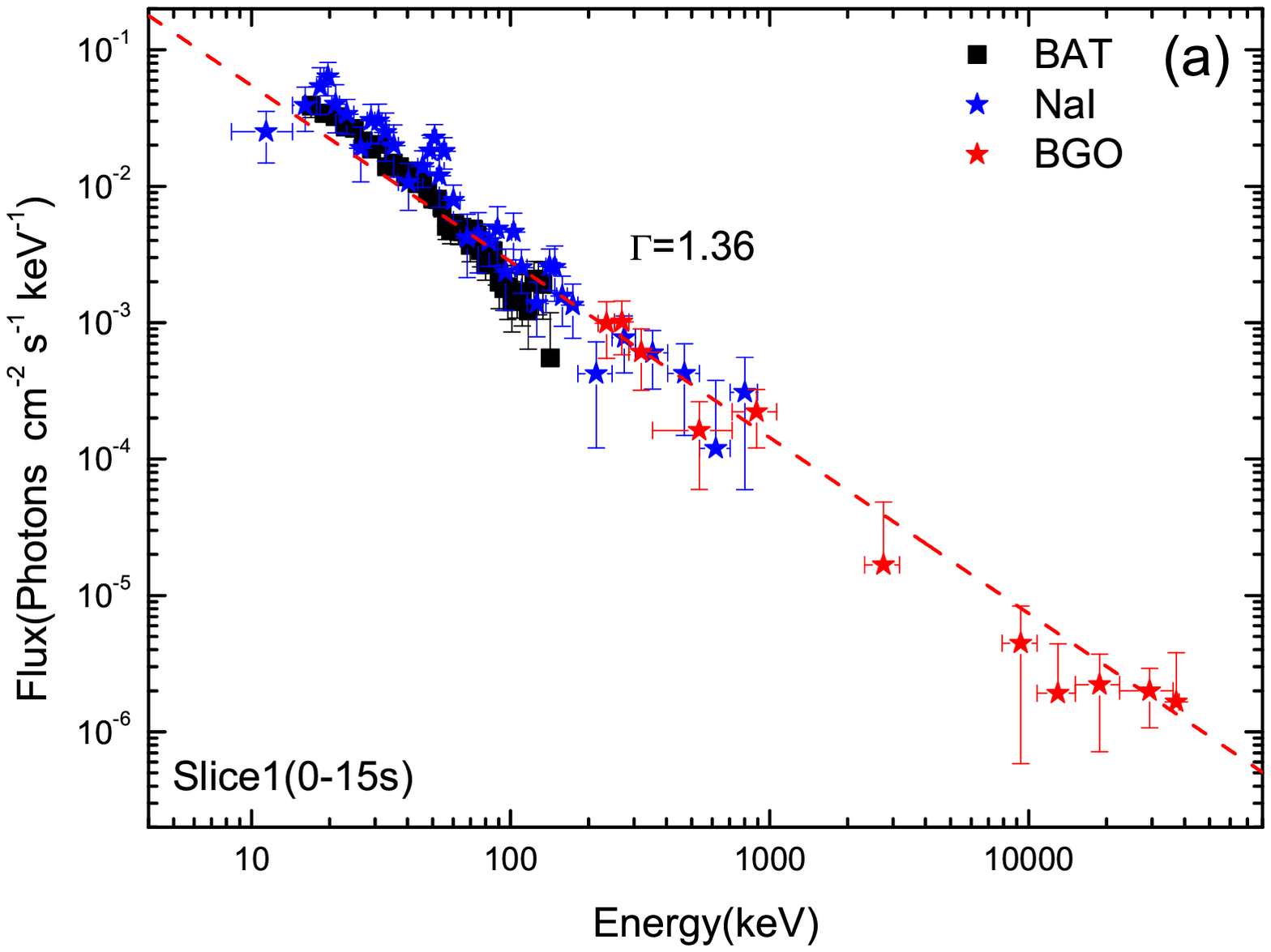}
\includegraphics[angle=0,width=0.5\textwidth]{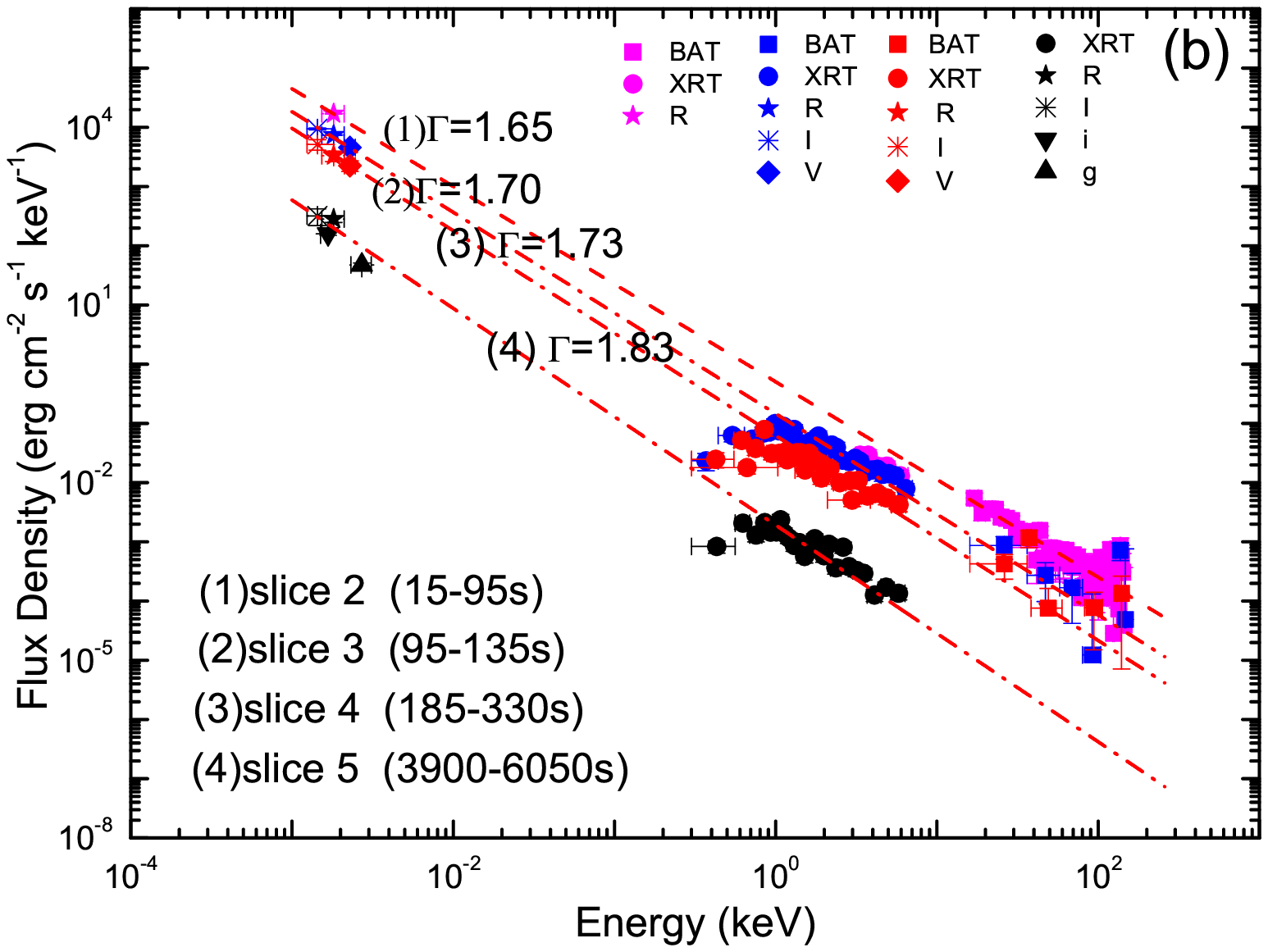}
\caption{The SED analysis of GRB 120729A. (a) Joint spectral fits of {\em Swift}/BAT and $Fermi$/GBM. They can be well fitted with a single power-law function, with photon index $\Gamma_{\rm \gamma}=1.36$ and  $\chi^{2}=2.13$. (b) Joint spectral fits of the BAT, X-ray, and optical afterglow in four selected time intervals. The dashed lines show the intrinsic
power-law spectra derived from the joint fits. The photon indices are also marked.}
\label{spectral_analysis}
\end{figure}

The fitting results are presented in Figure \ref{spectral_analysis} and Table \ref{tab-SED}. The SED of Slice 1 shows the BAT spectrum and GBM spectrum fitted with a single power-law function, with a photon index $\Gamma_{\rm \gamma}=1.36\pm0.02$ (seen in Figure \ref{spectral_analysis}a) and a quite large $\chi^{2}=2.13$. Calibration inconsistencies among the different instruments (GBM, BAT) could produce the high $\chi^{2}$. The SED of the joint $\gamma$-ray, X-ray, and optical spectrum (Slices 2, 3, and 4; see Figure \ref{spectral_analysis}b) can be well fitted with a single absorbed power-law function, without considering the host-galaxy extinction. Their corresponding photon indices are 1.65, 1.70, and 1.73 for (respectively) Slices 2, 3, and 4. The SED of the joint optical and X-ray spectrum at the late-time epoch (Slice 5, seen in Figure \ref{spectral_analysis}b) can also be well fitted with a single absorbed power-law function, with photon index 1.83. We take $\Gamma = 1.75$ as a rough average (and stable) value of the photon index during the afterglow phase.

Assuming the GRB 120729A multiwavelength data are located in the same spectral regime ($\nu_{\rm m}<\nu<\nu_{\rm c}$)  and the circumburst medium is just the interstellar medium (ISM), one has the $\alpha-\beta$ closure relation with $\alpha= 3\beta /2$, $\alpha= 3\beta /2+0.75$, and $\alpha=(q-1)+(2+q)\beta/2$ for the normal decay phase, the shallow decay phase, and the post-jet-break phases (respectively). Here the parameter $q$ is the energy injection parameter, which represents the central engine with a power-law luminosity history $L(t)=L_{0}(t/t_{0})^{-q}$ \citep[][]{zhangmeszaros01,zhang06}.

Combining the results of our temporal analysis and spectral analysis, we use the closure relation ($\alpha-\beta$) of the fireball external shock model to test the multiwavelength data. We can see that the rising slope of the smooth bump is $\alpha_{\rm BAT,I}=-0.86\pm0.11$ (Phase I). After the fireball starts decelerating, it transitions to a normal decay phase with a theoretical value  $\alpha= 3\beta /2=3/2\times0.75=1.13$. The average of the observational spectral indices after the deceleration time is $\beta=\Gamma - 1 \approx 0.75$, without considering the uncertainties. The observed values, $\alpha_{\rm BAT,II}=\alpha_{\rm X,II}=1.27\pm0.09$ and $\alpha_{\rm R,II}=0.96\pm0.02$ in Phase II, are approximately equivalent to the theoretical ones. In Phase III, we obtained an energy injection parameter $q=\frac{(2+2\alpha-2\beta)}{(2-\beta)}\approx 0.12$ for both the X-ray and optical afterglow with  $\alpha_{\rm X,III}=0.18$ (fixed) and $\alpha_{\rm R,III}=0.18\pm0.12$, which is located in a reasonable range predicted by the external shock. The theoretical value of the post-jet break is $\alpha= 3\beta /2+0.75=1.88$, which is also close to the observed one, $\alpha_{\rm X,IV}=1.83$ and $\alpha_{\rm R,IV}=1.70\pm0.09$ (Phase IV).

 The results of the temporal and spectral analyses show that the multiwavelength observed light curves have an achromatic behavior, consistent with the predictions of the external forward shock models with energy injection in the thin-shell case \citep{zhang06,lil12},
 suggesting that both the prompt $\gamma$-ray emission and the afterglow have the same physical origin.
 The rising slope of this model for the forward-shock onset is typically steep \citep{gao13}.
 On the other hand, the circumburst medium may be more complex earlier \citep[e.g.,][]{dai03,liang10,liang13,yi13},
 with the possibility of a transition between a wind profile and a constant-density medium. Adopting a more general
 profile with power-law index $k$ of the circumburst medium [$n(r)\propto r^{-k}$], the predicted temporal index
 is \textbf{$\alpha=3-[k(\beta+3)/2]$} \citep{liang13}. The rising slope of the smooth bump of GRB 120729A ($\alpha_{\rm BAT,I}=-0.86\pm0.11$)
  is consistent with a moderate $k$ value.

\section{MODELING: EXTERNAL SHOCK ORIGIN FOR BOTH THE PROMPT GAMMA-RAY EMISSION AND AFTERGLOW}

We further investigate our data with the standard forward shock model with energy injection. The light curves of GRB 120729A indicate that the $\gamma$-ray emission phase (Phase I) is the blast-wave deceleration phase.  It then transitions to the self-similar deceleration phase (Phase II), followed by the energy injection which may invoke either a long-lasting central engine \citep{dai98,zhangmeszaros01} or a Lorentz-factor-stratified ejecta \citep{rees98,sari00,uhm12} exhibiting a shallow decay (Phase III). A post-jet-break phase (Phase IV) is entered when the $1/\Gamma$ cone is no longer filled with emission owing to the edge effect, with a steepening light curve\citep[e.g.,][]{panaitescu98}. The late-time afterglow of GRB 120729A (beyond $10^5$~s) has a contribution from a possible accompanying supernova \citep{cano14} and is beyond the scope of this paper.

\begin{figure}[htbp]
\includegraphics[angle=0,width=0.5\textwidth]{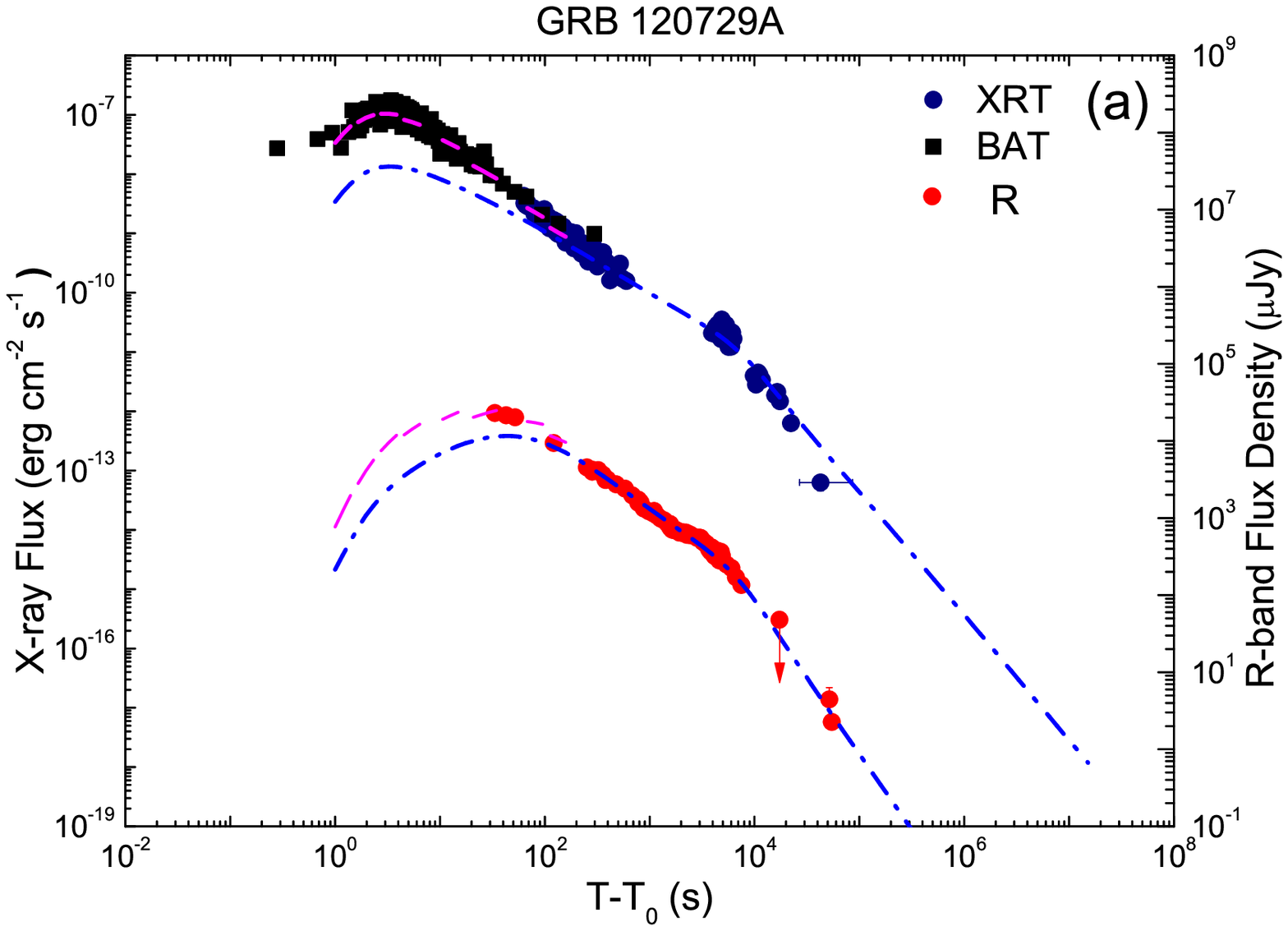}
\includegraphics[angle=0,width=0.45\textwidth]{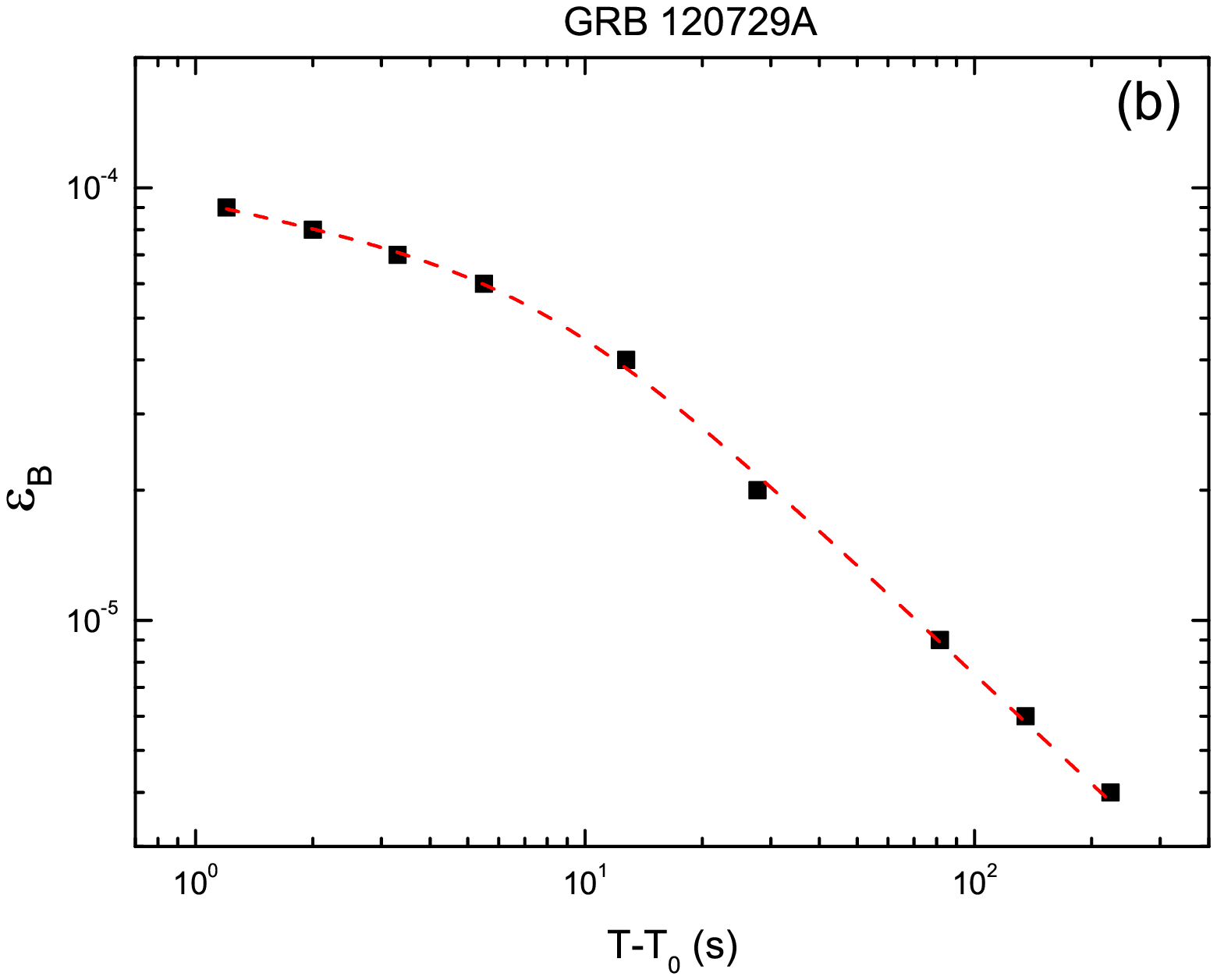}
\caption{Modeling the multiwavelength light curves with an external shock model by introduced  energy injection $L(t)=L_{0}(t/t_0)^{-q}$, and a time-dependent microphysics model with $\epsilon_B\propto t^{\alpha_B}$. (a) Comparisons of the modeling data and the observed data, the blue dashed line represents the constant microphysics model, while the pink dashed line represents the results of emission by considering a time-dependent microphysics model with $\epsilon_B\propto t^{\alpha_B}$ at early times, $T < T_{\rm 0} + 157$~s. (b) The value of $\epsilon_B$ during the early emission epoch, $T < T_{\rm 0} + 157$~s, fitted by a broken power-law function (dashed line), with $\alpha_{\rm B,1} = 0.18 \pm 0.04$, $\alpha_{\rm B,2} = 0.84 \pm 0.04$, and $t_{\rm \alpha_B,b} = 8.0 \pm 1.6$~s.}
\label{theory-fit}
\end{figure}

 We adopt the standard external shock model by Sari et al. (1998) and Huang et al. (1999). The spectra in both the $\gamma$-ray and afterglow regions are denoted as $N_e\propto \gamma_e^{-p}$. The spectral regimes are assumed to be located in $\nu_{\rm m}<\nu<\nu_{\rm c}$, and we fix $p=2\beta+1=2.5$. Our empirical analysis also indicates that the features of GRB 120729A emission are consistent with expectations in the ISM scenario. We adopt a constant ISM density $n$. The free parameters of our model include the isotropic kinetic energy $E_{\rm K,iso}$, the initial Lorentz factor $\Gamma_{\rm 0}$, the fraction of shock energy to electron energy $\epsilon_{\rm e}$, the fraction of shock energy to magnetic field energy $\epsilon_{\rm B}$, the jet opening angle $\theta_{\rm j}$, and the parameters of the energy injection $L_{\rm 0}$, $q$, $t_{\rm s}$ (energy injection starting time), and $t_{\rm e}$ (energy injection ending time). We can see that they can well constrain the observational data when $T > T_{\rm 0} + 157$~s (as shown in Figure \ref{theory-fit}a). When $T > T_{\rm 0} + 157$~s, the best-fitting parameters of the standard external forward shock model are $E_{\rm K,iso} = 3.36\times 10^{\rm 54}$ erg, $\Gamma_{\rm 0}=760$, $n = 8$ cm$^{\rm -3}$, $\epsilon_{\rm e} = 0.01$, $\epsilon_{\rm B} = 4.0\times 10^{\rm -6}$, $\theta_{\rm j} = 0.0238$ rad,  $L_{\rm 0} = 5.0\times10^{\rm 50}$ erg s$^{-1}$, $q = -0.1$, $t_{\rm s} = 0$~s, and $t_{\rm e} = 3698$~s.

In the standard model, the energy goes into electrons ($\epsilon_e$) and magnetic fields ($\epsilon_B$). The microphysical parameters are typically assumed to be not varying, and in fact constant $\epsilon_e$ and $\epsilon_B$ are consistent with the observations of late-time afterglows (e.g., Yost et al. 2003; Wang et al. 2015). The mechanism of energy transfer from protons to electrons and magnetic fields in the relativistic shocks is complicated. Modifications of the standard afterglow model with time-dependent microphysics models have been proposed to solve difficulties encountered with observations, such as X-ray afterglow plateaus, chromatic breaks, and afterglow rebrightenings (Ioka et al. 2006, Panaitescu et al. 2006; Kong et al. 2010; van der Horst et al. 2014). The early-time emission is even more complicated.

To explain the early emission of GRB 120729A, we consider a time-dependent microphysics model with $\epsilon_e\propto t^{\alpha_e}$ and/or $\epsilon_B\propto t^{\alpha_B}$ \citep{06ioka}. In the standard forward shock model, $\epsilon_e$ and $\epsilon_B$ evolve with $\epsilon_e \propto t^{-2p/(p+4)}$ and $\epsilon_B \propto t^{2p/(2p-1)}$ (for $\nu_{\rm m}<\nu<\nu_{\rm c}$ and $p>2$), respectively. Figure \ref{theory-fit}a shows that the modeled light curves with constant microphysics are lower than the observed data at early time ($T < T_{\rm 0} + 157$~s). We therefore assume that $\epsilon_B$ evolves with time and $\epsilon_e$ is still constant, while the rest of the parameters are the same as above. We found that modeling the light curves after modification is well consistent with the data for both the prompt $\gamma$-ray emission and the afterglow light curve (also shown in the Figure \ref{theory-fit}a). The value of $\epsilon_B$ in the early-emission epoch ($T < T_{\rm 0} + 157$~s) evolves from $2.0\times 10^{\rm -5}$ to $4.0\times 10^{\rm -6}$ (Figure \ref{theory-fit}b). It can be fitted with a broken power-law function, with $\alpha_{\rm B,1} = 0.18 \pm 0.04$, $\alpha_{\rm B,2} = 0.84 \pm 0.04$, and $t_{\rm \alpha_B,b} = 8.0 \pm 1.6$~s.

\section{Discussion and Conclusions}\label{sec:Conclusion}
 GRB 120729A has a single smooth $\gamma$-ray emission pulse detected by {\it Swift}/BAT and {\it Fermi}/GBM. Its broadband afterglow was detected by {\it Swift}/XRT, {\it Swift}/UVOT, and ground-based optical telescopes. We obtained well-sampled KAIT $BVRI$ light curves from 98.9~s to 1.4~hr after the {\it Swift}/BAT trigger. We found that the prompt $\gamma$-ray emission from {\it Swift}/BAT extends smoothly to the X-ray emission. Extensive analysis and modeling of the multiwavelength light curves shows that an external shock model can explain both the prompt $\gamma$-ray emission and the afterglow. The properties of GRB 120729A are summarized as follows.
 \begin{enumerate}
   \item The temporal and spectral joint fits of the multiwavelength light curves of GRB 120729A reveal achromatic behavior consistent with the external shock model. The light curves from the prompt $\gamma$-ray emission phase to the optical can be decomposed into four phases: onset bump (Phase I), normal decay (Phase II), shallow decay (Phase III), and post-jet break (Phase IV).
   \item There is no obvious evolution of the SED between the X-ray and optical afterglows, with an average value of the photon index $\Gamma \approx 1.75$. The SED exhibits slight evolution from the prompt $\gamma$-ray emission to the afterglow, with $\Gamma_{\rm \gamma}=1.36$ to $\Gamma \approx 1.83$.
   \item The multiwavelength light curves from the $\gamma$-ray emission to the optical afterglow can be well fitted with the external shock model, by introducing energy injection and time-dependent microphysics $\epsilon_B\propto t^{\beta_B}$.  The best parameters are $E_{\rm K,iso} = 3.36\times 10^{\rm 54}$ erg, $\Gamma_{\rm 0}=760$, $n = 8$ cm$^{\rm -3}$, $\epsilon_{\rm e} = 0.01$, $\epsilon_{\rm B} \approx [9\times 10^{\rm -5}, 4\times 10^{\rm -6}]$, $\theta_{\rm j} = 0.0238$ rad, $L_{\rm 0} = 5.0\times10^{\rm 50}$ erg s$^{\rm -1}$, $q = -0.1$, $t_{\rm s} = 0$~s, and $t_{\rm e} = 3698$~s. There is obvious evolution of $\epsilon_B$ from $9\times 10^{\rm -5}$ to $4\times 10^{\rm -6}$ in the early emission phase when $T < T_{\rm 0} + 157$~s.
 \end{enumerate}

 The theoretical model suggests that $\epsilon_B$ evolves with time as  $\epsilon_B \propto t^{2p/(2p-1)}$ for the ISM case with $\nu_{\rm m}<\nu<\nu_{\rm c}$ and $p>2$ \citep{06ioka}. For GRB 120729A, we can see that the photon index evolves from the prompt $\gamma$-ray emission to the afterglow with $\Gamma_{\rm \gamma}=1.36$ to $\Gamma \approx 1.75$. The $\epsilon_B$ in the early emission ($T < T_{\rm 0} + 157$~s) can be described as a broken power-law function, with $\alpha_{\rm B,1} = 0.18 \pm 0.04$, $\alpha_{\rm B,2} = 0.84 \pm 0.04$, and $t_{\rm \alpha_B,b} = 8.0 \pm 1.6$~s.  The results of the time-dependent microphysics model are consistent with the observed data. We obtain a low value of $\epsilon_B \approx [9\times 10^{\rm -5}, 4\times 10^{\rm -6}]$. The value of $\nu_c$ is very sensitive to $\epsilon_B$, with $\nu_c \propto \epsilon_B^{-3/2}$. The extremely low value of $\epsilon_B$ ensures that both the optical and X-ray emission are still in the regime $\nu_{\rm m}<\nu<\nu_{\rm c}$. These results are also consistent with previous studies \citep[e.g.,][]{kumar09,kumar10,santana14,duran14,wang15,zhong16}.

\begin{figure}[htbp]
\includegraphics[angle=0,width=0.5\textwidth]{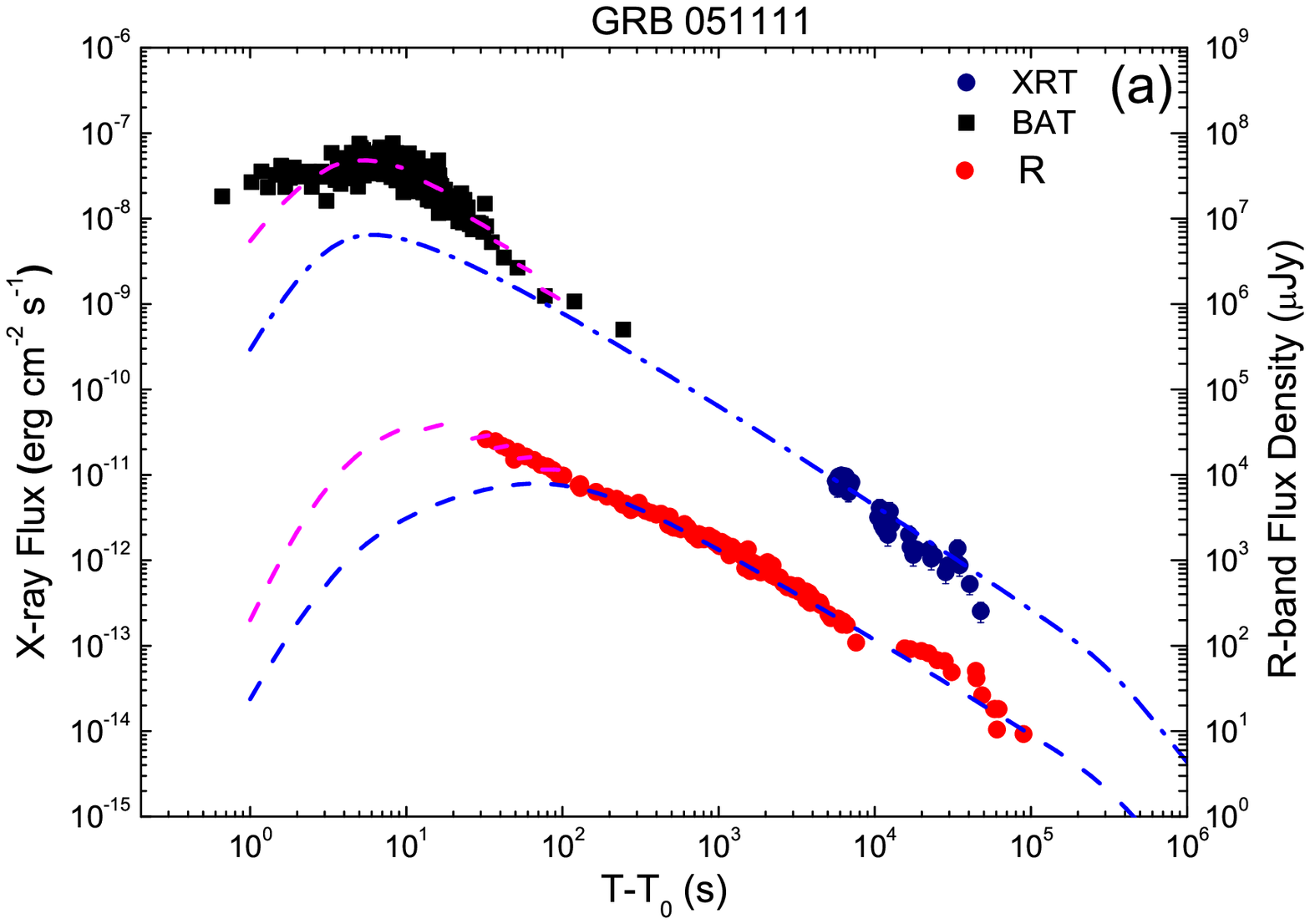}
\includegraphics[angle=0,width=0.45\textwidth]{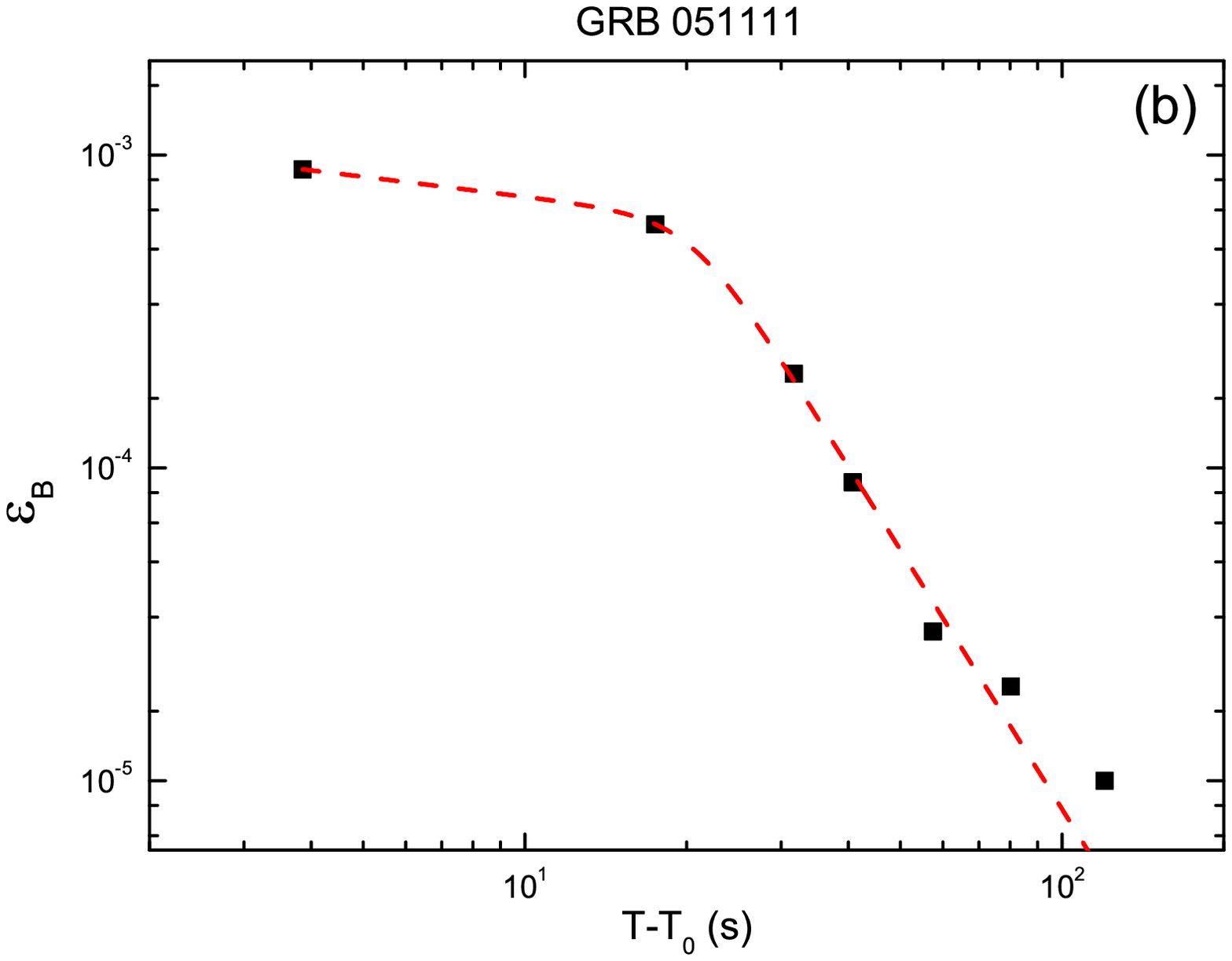}
\caption{The same as Figure \ref{theory-fit}, but for GRB 051111. (a) When $T < T_{\rm 0} + 120$~s, $\epsilon_B$ evolves as  $\epsilon_B\propto t^{\alpha_B}$, and then  $\epsilon_B$ stays constant after $T =T_{\rm 0} + 120$~s. (b) The value of $\epsilon_B$ during the early emission epoch, $T < T_{\rm 0} + 120$~s, fitted by a broken power-law function (dashed line), with $\alpha_{\rm B,1} = 0.21 \pm 0.08$, $\alpha_{\rm B,2} = 2.75 \pm 0.32$, and $t_{\rm \alpha_B,b} = 20.6 \pm 2.5$~s.}
\label{GRB-051111}
\end{figure}

\begin{figure}[htbp]
\includegraphics[angle=0,width=0.5\textwidth]{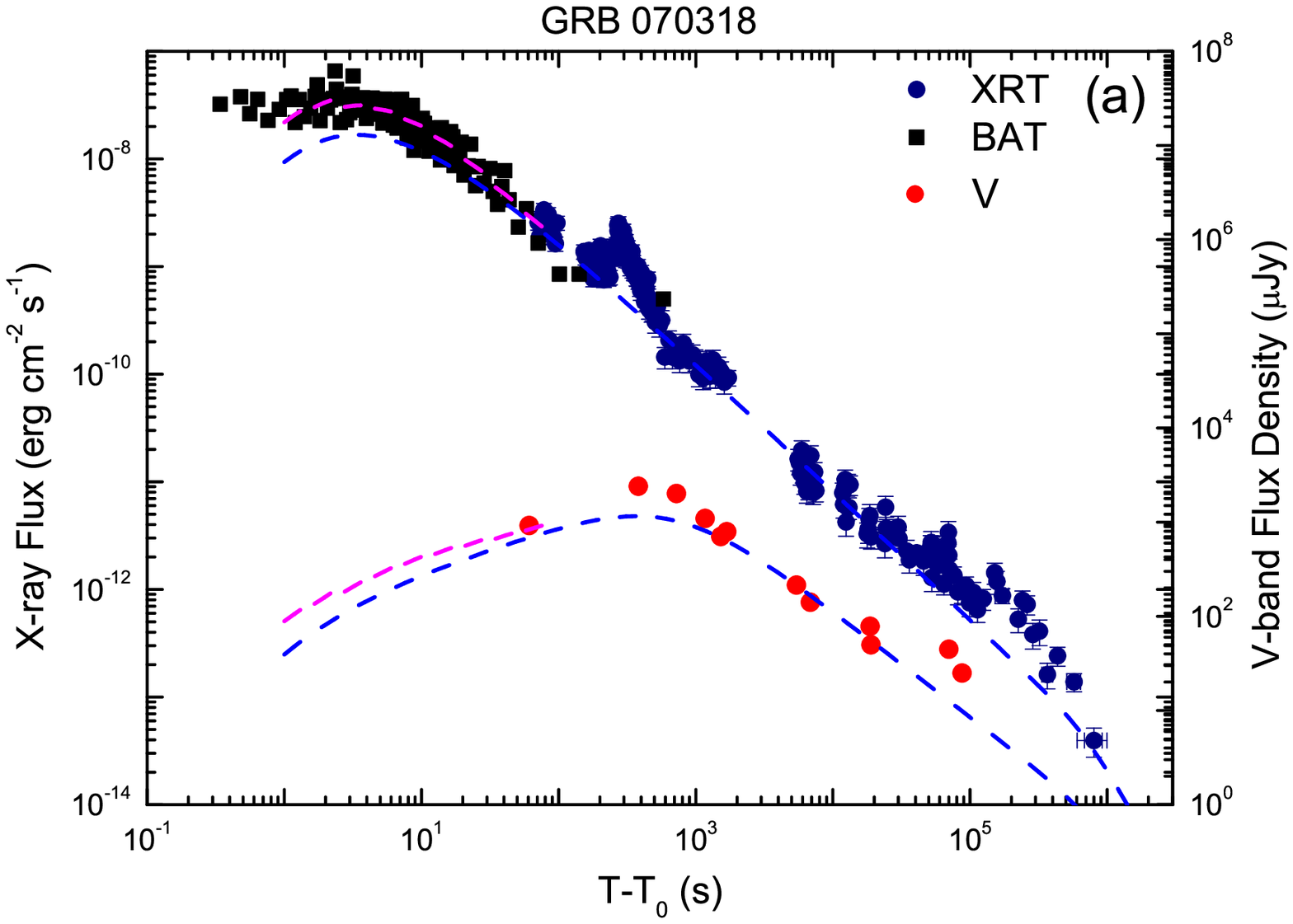}
\includegraphics[angle=0,width=0.45\textwidth]{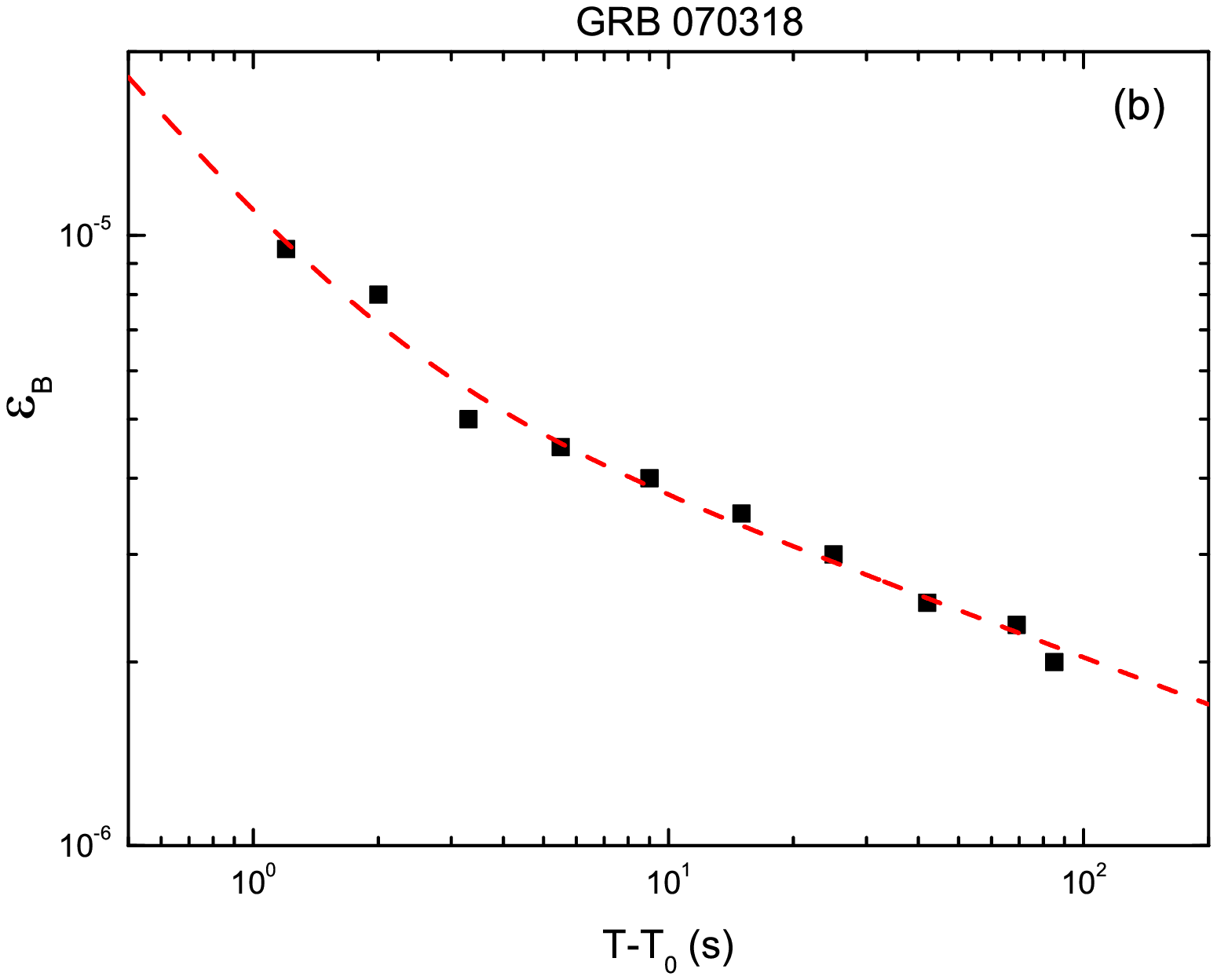}
\caption{The same as Figure \ref{theory-fit}, but for GRB 070318. (a) When $T < T_{\rm 0} + 85$~s, $\epsilon_B$ evolves as  $\epsilon_B\propto t^{\alpha_B}$, and then $\epsilon_B$ stays constant after $T =T_{\rm 0} + 85$~s. (b) The value of $\epsilon_B$ during the early emission epoch, $T < T_{\rm 0} + 85$~s, fitted by a broken power-law function (dashed line), with $\alpha_{\rm B,1} = 0.80 \pm 0.36$, $\alpha_{\rm B,2} = 0.25\pm 0.08$, and $t_{\rm \alpha_B,b} = 2.2\pm 2.0$~s.}
\label{GRB-070318}
\end{figure}

In general, the characteristics of external shock pulses (e.g., pulse width, peak energy in the GRB spectrum $E_p$, and spectral-index evolution with time) can be very different from those of internal dissipation pulses \citep[e.g.,][]{fishman95,norris96,peng07}. In order to systematically search for possible external-shock-origin GRBs, we collected candidate GRBs with a clear single pulse observed with {\it Swift}/BAT up to December 2017. There are 34 GRBs showing a very smooth single pulse, 29 a smooth single pulse followed by a small pulse, and 15 a small pulse followed by a smooth single pulse. The other 923 GRBs exhibit more pulses and are highly variable. We found that 11 GRBs in the single-pulse sample (GRBs 050721A, 050801, 051109A, 051111, 060912A, 070318, 070531, 080805, 110503A, 121117A, and 151006A) show a smooth connection between the BAT and XRT data with a slope less than the typical value of 1.5. Among them, 6 GRBs (GRBs 050721, 060912A, 070531, 080805, 121117A, and 151006A) lack early-time optical detections. Inspecting the other 5 GRBs, we find that only GRB 051111 and GRB 070318 may originate from an external shock for both the prompt $\gamma$-ray emission and the afterglow (as shown in Figures \ref{GRB-051111} and \ref{GRB-070318}). Similar to GRB 120729A, the BAT detections of GRB 05111 and GRB 070318 start 3.03~s and 0.28~s (respectively) before the BAT trigger; thus, we have  shifted $T_0$ to be the time of initial BAT detection.

For GRB 051111, the best-fit parameters are $E_{\rm K,iso} = 4.06\times 10^{\rm 54}$ erg, $\Gamma_{\rm 0}=650$, $n = 12$ cm$^{\rm -3}$, $\epsilon_{\rm e} = 0.01$, and $\theta_{\rm j} = 0.1$ rad. The $\epsilon_B$ value evolves as $\epsilon_B \propto t^{\alpha_B}$ at early times ($T < T_{\rm 0} + 120$~s) from $9.0\times 10^{\rm -4}$ to $1.0\times 10^{\rm -5}$; it can be fitted with a broken power-law function (dashed line) with $\alpha_{\rm B,1} = 0.21 \pm 0.08$, $\alpha_{\rm B,2} = 2.75 \pm 0.32$, and $t_{\rm \alpha_B,b} = 20.6 \pm 2.5$~s. Subsequently, $\epsilon_B$ stays constant at $\epsilon_{\rm B} = 1.0\times 10^{\rm -5}$. For GRB 070318, the best-fit parameters are $E_{\rm K,iso} = 3.86\times 10^{\rm 53}$ erg, $\Gamma_{\rm 0}=1000$, $n = 8$ cm$^{\rm -3}$, $\epsilon_{\rm e} = 0.12$, and $\theta_{\rm j} = 0.2$. The value of $\epsilon_B$ evolves as  $\epsilon_B\propto t^{\alpha_B}$ at early times ($T < T_{\rm 0} + 85$~s) from $9.5\times 10^{-6}$ to $2.0\times 10^{-6}$, with $\alpha_{\rm B,1} = 0.80 \pm 0.36$, $\alpha_{\rm B,2} = 0.25 \pm 0.08$, and $t_{\rm \alpha_B,b} = 2.2 \pm 2.0$~s. Thereafter ($T = T_{\rm 0} + 85$~s), $\epsilon_B$ stays constant at $\epsilon_{\rm B} = 2.0\times 10^{\rm -6}$.

No other GRBs display a smooth connection between the BAT and XRT data. In principle, all GRBs should have an external shock onset component similar to the one seen in GRB 120729A. However, whether it will appear in $\gamma$-rays depends on the shock parameters. Our results suggest that once internal dissipation occurs, the $\gamma$-ray efficiency is much higher than that of the external shock emission, and the external shock component is likely masked by internal dissipation emission in most events. This is consistent with theoretical modeling \citep{maxham09}.
It seems that only a small fraction of GRBs may originate from an external shock for both the prompt $\gamma$-ray emission and the afterglow.

If the composition of the GRB jet is a matter-dominated fireball, the absence of a ``prompt emission'' signal caused by internal dissipation within the jet is puzzling. Even though the lack of the internal shock emission may be circumvented by assuming that there is no significant variability within the outflow, the photospheric emission from the fireball would nevertheless show up, which should be above the detection threshold owing to the high efficiency expected for photospheric emission \citep{meszarosree00}. The absence of such bright emission before the external-shock-origin $\gamma$-ray peak therefore suggests that the outflow is Poynting-flux dominated, so that the photospheric luminosity is suppressed by a factor of $(1+\sigma_{\rm ph})$, where $\sigma_{\rm ph}$ is the magnetization parameter at the photosphere \citep{zhangpeer09,15gao}. Such an outflow may still keep a moderate $\sigma$ at the deceleration radius, so that the reverse-shock emission is also suppressed \citep{zhangk05,gao15b}. Incidentally, the three GRBs we are studying (GRBs 051111, 070318, and 120729A) all did not show a reverse-shock emission component in the optical band, suggesting a self-consistent picture. It is possible that a dissipationless Poynting-flux-dominated outflow is a requirement to produce external-shock-origin single $\gamma$-ray pulses. The scarcity of such bursts is consistent with the hypothesis that GRBs have diverse jet compositions \citep{zhang11}, and the pure fireballs and pure Poynting-flux-dominated flows reflect the two extreme regimes of the jet composition.

\acknowledgments
We thank an anonymous referee for helpful suggestions. This work is supported by the National Basic Research Program of China (973 Program, Grant No. 2014CB845800),
the National Natural Science Foundation of China (Grant No. 11673006, 11533003), and the Guangxi Science
 Foundation (Grant No. 2016GXNSFFA380006, AD17129006).  We also acknowledge the use of public data from the \emph{Swift} and {\it Fermi} data archives.
A.V.F.'s supernova/GRB group is grateful for financial
assistance from NSF grant AST-1211916, the TABASGO Foundation,
the Christopher R. Redlich Fund, NASA/{\it Swift} grants NNX10AI21G
and NNX12AD73G, and the Miller Institute for Basic
Research in Science (U.C. Berkeley).
KAIT and its ongoing operation were made possible by
donations from Sun Microsystems, Inc., the Hewlett-Packard Company,
AutoScope Corporation, Lick Observatory, the NSF, the University of
California, the Sylvia \& Jim Katzman Foundation, and the TABASGO
Foundation.
Research at Lick Observatory is partially supported by a generous
gift from Google.



\clearpage
\begin{deluxetable}{cccccc}

\tabletypesize{\small}
\tablecaption{KAIT Observations of GRB 120729A}
\tablewidth{0pt}

\tablehead{
\colhead{$T-T_{\rm 0}$(mid) (s)$^{\rm a}$}&
\colhead{Exp. (s)}&
\colhead{Mag$^{\rm b}$}&
\colhead{$\sigma^{\rm c}$} &
\colhead{Filter}&
}
\startdata
\hline\hline
101.98	&	20	&14.7	&	0.04	&	V	\\
202.18	&	20	&15.3	&	0.04	&	V	\\
299.88	&	20	&15.7	&	0.04	&	V	\\
400.08	&	20	&15.9	&	0.04	&	V	\\
500.28	&	20	&16.0	&	0.04	&	V	\\
598.78	&	20	&16.2	&	0.05	&	V	\\
698.98	&	20	&16.4	&	0.06	&	V	\\
799.18	&	20	&16.7	&	0.07	&	V	\\
899.48	&	20	&16.8	&	0.08	&	V	\\
996.18	&	20	&16.9	&	0.09	&	V	\\
1093.88	&	20	&16.9	&	0.09	&	V	\\
1194.08	&	20	&17.0	&	0.11	&	V	\\
1294.28	&	20	&17.1	&	0.1	&	V	\\
1394.48	&	20	&17.3	&	0.1	&	V	\\
1492.98	&	20	&17.2	&	0.12	&	V	\\
134.78	&	20	&13.8	&	0.03	&	I	\\
235.08	&	20	&14.3	&	0.04	&	I	\\
333.48	&	20	&14.6	&	0.04	&	I	\\
432.88	&	20	&14.8	&	0.04	&	I	\\
533.08	&	20	&14.9	&	0.05	&	I	\\
632.48	&	20	&15.1	&	0.05	&	I	\\
731.88	&	20	&15.4	&	0.05	&	I	\\
832.08	&	20	&15.6	&	0.07	&	I	\\
930.58	&	20	&15.7	&	0.06	&	I	\\
1028.18	&	20	&15.8	&	0.06	&	I	\\
1127.58	&	20	&15.9	&	0.07	&	I	\\
1226.88	&	20	&16.0	&	0.07	&	I	\\
1327.08	&	20	&16.1	&	0.08	&	I	\\
1427.38	&	20	&16.1	&	0.08	&	I	\\
1525.88	&	20	&16.2	&	0.09	&	I	\\
1593.28	&	20	&16.4	&	0.1	&	I	\\
1659.78	&	20	&16.4	&	0.12	&	I	\\
1723.68	&	20	&16.5	&	0.12	&	I	\\
1797.18	&	20	&16.4	&	0.09	&	I	\\
1866.28	&	20	&16.5	&	0.11	&	I	\\
1937.08	&	20	&16.4	&	0.09	&	I	\\
2007.98	&	20	&16.6	&	0.13	&	I	\\
2078.78	&	20	&16.5	&	0.12	&	I	\\
2150.48	&	20	&16.6	&	0.13	&	I	\\
2220.48	&	20	&16.6	&	0.12	&	I	\\
2291.38	&	20	&16.6	&	0.12	&	I	\\
2362.18	&	20	&16.6	&	0.12	&	I	\\
2433.08	&	20	&16.6	&	0.11	&	I	\\
2503.88	&	20	&16.6	&	0.11	&	I	\\
2884.08	&	20	&16.7	&	0.11	&	I	\\
2954.88	&	20	&16.7	&	0.13	&	I	\\
3025.78	&	20	&16.8	&	0.13	&	I	\\
3097.48	&	20	&16.8	&	0.12	&	I	\\
3168.28	&	20	&16.8	&	0.13	&	I	\\
3239.18	&	20	&16.9	&	0.12	&	I	\\
3431.88	&	20	&17.0	&	0.14	&	I	\\
3505.28	&	20	&17.0	&	0.13	&	I	\\
3576.08	&	20	&17.1	&	0.14	&	I	\\
3648.68	&	20	&17.2	&	0.16	&	I	\\
3719.58	&	20	&16.9	&	0.14	&	I	\\
3790.38	&	20	&17.1	&	0.14	&	I	\\
3861.28	&	20	&17.0	&	0.12	&	I	\\
3932.08	&	20	&17.2	&	0.18	&	I	\\
4002.98	&	20	&17.2	&	0.17	&	I	\\
4073.78	&	20	&17.1	&	0.15	&	I	\\
4145.48	&	20	&17.2	&	0.18	&	I	\\
4216.38	&	20	&17.5	&	0.22	&	I	\\
4287.18	&	20	&17.2	&	0.16	&	I	\\
4357.18	&	20	&17.2	&	0.17	&	I	\\
4427.98	&	20	&17.3	&	0.19	&	I	\\
4498.88	&	20	&17.4	&	0.22	&	I	\\
4569.68	&	20	&17.3	&	0.21	&	I	\\
4641.48	&	20	&17.6	&	0.32	&	I	\\
4712.28	&	20	&17.3	&	0.23	&	I	\\
4783.08	&	20	&17.6	&	0.36	&	I	\\
4853.98	&	20	&17.7	&	0.31	&	I	\\
4924.78	&	20	&17.4	&	0.3	&	I	\\
4996.58	&	20	&17.5	&	0.3	&	I	\\
5067.38	&	20	&17.4	&	0.35	&	I	\\
5208.18	&	20	&17.4	&	0.45	&	I	\\
283.08	&	20	&15.1	&	0.01	&	R	\\
380.08	&	20	&15.4	&	0.01	&	R	\\
480.08	&	20	&15.5	&	0.01	&	R	\\
580.08	&	20	&15.7	&	0.01	&	R	\\
680.08	&	20	&15.9	&	0.01	&	R	\\
779.08	&	20	&16.1	&	0.01	&	R	\\
879.08	&	20	&16.3	&	0.01	&	R	\\
977.08	&	20	&16.4	&	0.01	&	R	\\
1075.08	&	20	&16.4	&	0.01	&	R	\\
1175.08	&	20	&16.5	&	0.01	&	R	\\
1274.08	&	20	&16.6	&	0.01	&	R	\\
1374.08	&	20	&16.7	&	0.01	&	R	\\
1474.08	&	20	&16.8	&	0.01	&	R	\\
1574.08	&	20	&16.9	&	0.01	&	R	\\
1640.08	&	20	&17.0	&	0.01	&	R	\\
1705.08	&	20	&17.0	&	0.01	&	R	\\
1778.08	&	20	&17.0	&	0.01	&	R	\\
1844.08	&	20	&17.0	&	0.01	&	R	\\
1915.08	&	20	&17.1	&	0.01	&	R	\\
1986.08	&	20	&17.1	&	0.01	&	R	\\
2057.08	&	20	&17.1	&	0.01	&	R	\\
2128.08	&	20	&17.1	&	0.01	&	R	\\
2199.08	&	20	&17.1	&	0.01	&	R	\\
2270.08	&	20	&17.2	&	0.01	&	R	\\
2341.08	&	20	&17.1	&	0.01	&	R	\\
2412.08	&	20	&17.2	&	0.01	&	R	\\
2483.08	&	20	&17.2	&	0.01	&	R	\\
2933.08	&	20	&17.3	&	0.01	&	R	\\
3004.08	&	20	&17.3	&	0.01	&	R	\\
3075.08	&	20	&17.3	&	0.01	&	R	\\
3146.08	&	20	&17.3	&	0.01	&	R	\\
3217.08	&	20	&17.4	&	0.01	&	R	\\
3288.08	&	20	&17.4	&	0.01	&	R	\\
3483.08	&	20	&17.4	&	0.01	&	R	\\
3554.08	&	20	&17.5	&	0.01	&	R	\\
3625.08	&	20	&17.6	&	0.01	&	R	\\
3698.08	&	20	&17.6	&	0.01	&	R	\\
3769.08	&	20	&17.6	&	0.01	&	R	\\
3840.08	&	20	&17.7	&	0.01	&	R	\\
3911.08	&	20	&17.6	&	0.01	&	R	\\
3982.08	&	20	&17.6	&	0.01	&	R	\\
4053.08	&	20	&17.8	&	0.01	&	R	\\
4123.08	&	20	&17.8	&	0.01	&	R	\\
4194.08	&	20	&17.9	&	0.01	&	R	\\
4265.08	&	20	&17.8	&	0.01	&	R	\\
4336.08	&	20	&17.9	&	0.01	&	R	\\
4407.08	&	20	&17.7	&	0.01	&	R	\\
4478.08	&	20	&17.8	&	0.01	&	R	\\
4549.08	&	20	&17.9	&	0.01	&	R	\\
4620.08	&	20	&18.0	&	0.01	&	R	\\
4761.08	&	20	&17.7	&	0.01	&	R	\\
4832.08	&	20	&17.8	&	0.01	&	R	\\
4903.08	&	20	&17.9	&	0.01	&	R	\\

\hline
\enddata
\tablenotetext{a}{$T-T_{\rm 0}$ is the midpoint of each observation.
The reference time $T_{\rm 0}$ is the time of initial BAT detection,
 which is 3.08~s before the BAT trigger.}
\tablenotetext{b}{Not taking into account the Galactic extinction.}
\tablenotetext{c}{$\sigma$ is the uncertainty in the magnitude.}
\label{table:magnitude}
\end{deluxetable}

\clearpage
\begin{deluxetable}{ccccccccccccccccccccccccc}
\tabletypesize{\tiny}
\tablecaption{Temporal Analysis of GRB 120729A from $\gamma$-rays to Afterglow. }
\tablewidth{0pt}
\tabcolsep=5.5pt
\tablehead{ \colhead{Band}&
\colhead{$F_{\rm 0}$\tablenotemark{a}}&
\colhead{$\alpha_{\rm I}$}&
\colhead{$\alpha_{\rm II}$}&
\colhead{$t_{\rm b,1}$\tablenotemark{b}}&
\colhead{$F_{\rm 1} $\tablenotemark{c}}&
\colhead{$\alpha_{\rm III}$}&
\colhead{$\alpha_{\rm IV}$}&
\colhead{$t_{\rm b,2}$\tablenotemark{b}}
}
\startdata
  BAT    & $(1.60\pm0.04)\times10^{-7}$       & $-0.86\pm0.11$   & $1.27\pm0.09$      & $3.52\pm0.18$ &  $-$       & $-$   & $-$      & $-$    \\
  XRT    &  $-$       & $-$   & $1.27\pm0.09$      & $-$ &  $4.6\times10^{-11} ({\rm fixed})$       & $0.18 ({\rm fixed})$   & $1.83\pm0.08$      & $3153 ({\rm fixed})$    \\
  $R$    &  $-$       & $-$   & $0.91\pm0.01$      & $-$  &  $632.16\pm53.13$       & $0.18\pm0.12$   & $1.70\pm0.09$      & $3153\pm262$  \\
\enddata
\tablenotetext{a}{Flux at break time of the BAT data, in units of $\rm erg~ cm^{-2}~ s^{-1}$.}
\tablenotetext{b}{Break time, in units of seconds.}
\tablenotetext{c}{Flux at break time, in units of $\rm erg~cm^{-2}~s^{-1}$ and  $\rm uJy$ for the X-ray and $R$-band data, respectively.}
\label{table:temperol}
\end{deluxetable}

\begin{deluxetable}{ccccccccccccccccccccccccc}
\tablecaption{Spectral Analysis of $\gamma$-rays and Afterglow.}
\tablewidth{0pt}
\tablehead{ \colhead{Slice}&
\colhead{Interval (s)}&
\colhead{$\Gamma$}&
\colhead{$\chi_{\rm r}^{\rm 2}$}}
\startdata
  1   &    0--15  &  $1.36\pm0.02$ &  $2.13$ \\
  2   &    15--95   &  $1.65\pm0.01$ &  $0.97$\\
  3   &    95--135  &  $1.70\pm0.01$   &  $1.96$ \\
  4   &  185--330    &  $1.73\pm0.02$  & $1.29$  \\
  5   &    3900--6050   &   $1.83\pm0.02$  & $1.99$ \\
\enddata
\tablecomments{The hydrogen column density of the Milky Way is fixed at $1.0\times10^{\rm 21}~{\rm cm}^{-2}$. Optical extinction and neutral hydrogen absorption of soft X-rays in the GRB host galaxy are taken into account, but they are negligible.}
\label{tab-SED}
\end{deluxetable}

\clearpage

\end{document}